\documentclass[11pt]{article}
\usepackage[margin=1in]{geometry}
\usepackage{amsmath,amssymb,amsthm,mathtools,booktabs}
\usepackage[numbers,sort&compress]{natbib}
\usepackage{graphicx}
\usepackage{tikz}
\usepackage{quantikz}
\usepackage{xcolor}
\usepackage{caption}
\usepackage{subcaption}
\usepackage{listings}
\usepackage[hidelinks]{hyperref}

\theoremstyle{definition}
\newtheorem{theorem}{Theorem}
\newtheorem{proposition}[theorem]{Proposition}
\newtheorem{lemma}[theorem]{Lemma}

\theoremstyle{definition}
\newtheorem{remark}[theorem]{Remark}

\providecommand{\ket}[1]{\lvert #1\rangle}
\providecommand{\bra}[1]{\langle #1\rvert}
\newcommand{\E}{\mathbb{E}}

\newcommand{\tr}{\operatorname{tr}}
\newcommand{\Cov}{\operatorname{Cov}}
\newcommand{\one}{\mathbf 1}
\newcommand{\W}{\mathcal W}
\newcommand{\A}{\mathcal A}

\title{Hurst Exponent Estimation with Quantum Wavelets}
\author{Brani Vidakovic\\[2pt]
\small Department of Statistics, Texas A\&M University}
\date{}

\begin{document}
\maketitle

\begin{abstract}
How accurately can the Hurst exponent $H$ of fractional Brownian motion be estimated from repeated preparations of a single amplitude-encoded path, when every measurement interrogates the same realization? We extend the $\epsilon$-decimated quantum nondecimated wavelet transform in three directions: we remove the endpoint jump of circular filtering, derive the sampling distribution of the estimator, and compare readouts at equal query budgets. A one-qubit mirror extension removes the jump and keeps every coefficient that straddles a fold of the same order as an interior coefficient, for any compactly supported filter and circular shift. The transform for any such filter is compiled from the paraunitary lattice of its filter bank, and the energy share of a level is the success probability of one multiply-controlled reflection, which is both the controlled unitary of a Hadamard test and the oracle of amplitude estimation. For filters with at least two vanishing moments we prove a root-$N$ central limit theorem for the log-energy contrasts, independent of the limiting path norm, so the studentized estimator is asymptotically standard normal for deterministic query budgets under which the success count at every level diverges. For sparse counts we give a working multinomial--lognormal likelihood, exact in its readout model for direct measurement, and check its coverage by simulation. For paths of length $512$, mirroring reduces RMSE by a factor of $3$--$10$ for $H\ge0.5$, and with exact shares the nondecimated db2 estimator is $6$--$14\%$ above the root Cram\'er--Rao bound. Hadamard level tests are less accurate than direct measurement of the leading qubits at every budget. At $H=0.9$ and $2^{20}$ queries, maximum-likelihood amplitude estimation attains RMSE $0.0347$, against $0.0451$ for direct measurement and $0.0295$ with exact shares.
\end{abstract}


\section{Introduction}\label{sec:intro}

How accurately can the Hurst exponent $H$ of fractional Brownian motion (fBm) be estimated from repeated preparations of a single amplitude-encoded path, and which readout uses a finite budget of preparations best? Every measurement of such a state interrogates the same realization, so the error has two sources, the record and the shots, and the shot noise depends on the record through its norm.

The Hurst exponent governs the distribution of energy across scales. In a wavelet decomposition, the expected total detail energy per level grows asymptotically by a factor of $2^{2H}$ per step toward coarser levels, combining coefficient variance and the number of coefficients \citep{Flandrin1992,AbryVeitch1998}. An orthonormal wavelet transform can be implemented as a quantum unitary \citep{Hoyer1997,FijanyWilliams1998,Klappenecker1999}, now for wavelets of any order \citep{BagherimehrabAspuruGuzik2024}, and sample paths of fBm can themselves be prepared as amplitude-encoded states with gate complexity polylogarithmic in the number of time steps \citep{BoulandDandapaniPrakash2023}. Applied to an amplitude-encoded path, it makes the probability of measuring a level equal to the share of energy in that level. The $\epsilon$-decimated quantum nondecimated wavelet transform (QNDWT) of \citet{Vidakovic2025QNDWT} builds the whole family of circularly shifted orthogonal transforms coherently, with a shift register in uniform superposition, controlled circular shifts and one wavelet unitary; tracing out the register gives shift-averaged level energies, expectations of level projectors recover them, and their log-spectrum carries $H$. That construction leaves the inferential question open, and this paper advances it in three directions: it removes the endpoint jump that circular filtering creates in an fBm path, derives the sampling distribution of the resulting estimator of $H$ under finite query budgets, and compares readouts of the level shares at equal budgets.

Prior work does not answer the question. Quantum wavelet circuits act on periodic sequences \citep{Hoyer1997,FijanyWilliams1998,Klappenecker1999,BagherimehrabAspuruGuzik2024}, and the $\epsilon$-decimated QNDWT inherits their circular boundary, which for an fBm path joins the two endpoints through a jump; the classical limit theory of wavelet variances of fBm assumes exactly computed coefficients and boundaries handled by discarding coefficients \citep{Flandrin1992,IstasLang1997,Coeurjolly2001,Bardet2002}, which has no direct quantum counterpart, and it has no notion of a finite number of measurements of a single state.

Periodic filtering joins the two endpoints of the path, producing a jump with root mean square of order $n^H$ for a record of length $n$. A few coefficients can then account for most of the energy at a fine level when $H>1/2$ (Figure~\ref{fig:boundary}). The mirror extension in Section~\ref{sec:mirror} removes this jump before filtering. It uses one extra qubit and bounds the second moment of every zero-sum filter translate, including translates crossing a fold. Fold coefficients can have a different covariance from interior coefficients; the finite-record calibration accounts for this difference, and Theorem~\ref{thm:clt} shows that their contribution vanishes at the $N^{-1/2}$ scale.

Section~\ref{sec:lattice} compiles the periodic transform using the paraunitary lattice of a compactly supported orthonormal two-channel filter bank \citep{VaidyanathanHoang1988,Vaidyanathan1993}. Each rotation acts on the channel qubit, which distinguishes even and odd samples, and each delay, a one-step circular shift of the odd samples, is implemented as a decrement of the position register controlled on the channel qubit, that is, as a permutation. The construction allows the filter to be chosen for its statistical properties, and the shift-register construction in Section~\ref{sec:shift} produces nondecimated level shares. Figure~\ref{fig:pipeline} shows the complete circuit.

We reuse the shift-register construction of \citet{Vidakovic2025QNDWT} and build on it results of two kinds. On the quantum side, a one-qubit mirror extension makes circular filtering of a path free of boundary jumps for every compactly supported filter and every circular shift, a lattice compilation gives the transform for any such filter from single-qubit rotations of the channel qubit and controlled permutations of the position register, and a single multiply-controlled reflection, independent of the filter, gives the level share both as the success probability of a Hadamard test and as the target amplitude of amplitude estimation. On the statistical side, we prove a root-$N$ central limit theorem for the mirrored log-energy contrasts with asymptotically negligible fold terms (Theorem~\ref{thm:clt}), and we show that the fine-scale fluctuations are asymptotically independent of the path norm (Lemma~\ref{lem:indep}). The second result is what single-record quantum inference needs: the path norm sets the shot-noise variance and the GLS weights, so a deterministic query budget that grows with the record yields a variance mixture of normals, and the independence makes the studentized estimator asymptotically standard normal nonetheless. For sparse counts, integrating the normalization in closed form gives a working likelihood, a multinomial--lognormal mixture over the $K-1$ contrast dimensions whose multinomial readout model is exact for direct measurement and which accepts levels without successes; its calibration is checked by simulation.

Finally, on the same records and at equal query budgets, we compare direct measurement of the leading qubits, Hadamard level tests, and maximum-likelihood amplitude estimation, whose oracle is the level reflection. Direct measurement is more efficient than level tests at every setting, because each of its shots informs every level, and amplitude estimation is the most accurate when the informative shares are rare: at $H=0.9$ and $2^{20}$ queries its RMSE is $23\%$ below that of direct readout on the same records. At the smallest budget it returns no estimate on $16\%$ of records at $H=0.9$, and its advantage there is conditional on success. Its ideal CNOT cost per query is within $2\%$ of sampling, but its circuits are deep: at $2^{20}$ queries the longest one applies $256$ Grover iterates coherently, about $2.6\times10^6$ CNOT gates excluding record preparation, against about $5\times10^3$ for one shot of direct readout.

\section{Setting}\label{sec:setting}

The record, its encoding, the indexing of levels and the access model are fixed here.

\paragraph{Record.}
The record is a sampled fBm path $B=(B_0,\dots,B_{n-1})$, $B_t=\sum_{s\le t}x_s$, where $x$ is fractional Gaussian noise with variance $\sigma^2$ and $0<H<1$. After mirror extension (Section~\ref{sec:mirror}) the transformed vector $\tilde B$ has length $N=2n=2^J$.

\paragraph{Multiresolution indexing.}
A vector of length $N=2^J$ is an element of $V_J$. A decomposition with $L$ levels is
$$
V_J=V_{J-L}\oplus W_{J-L}\oplus W_{J-L+1}\oplus\cdots\oplus W_{J-1},
$$
where $W_j$ holds the $2^j$ detail coefficients of level $j$. The finest level of detail is $J-1$, the coarsest is $J-L$, and $V_{J-L}$ is the smooth space; the index $j$ decreases as resolution decreases. In Mallat order $[s_{J-L}\mid d_{J-L}\mid\cdots\mid d_{J-1}]$, level $j$ occupies the indices $[2^j,2^{j+1})$. A \emph{window} $\W$ of $K$ fine levels is $\{J-K,\dots,J-1\}$.

\paragraph{Encoding.}
The state is $\ket{y}=\|y\|^{-1}\sum_ty_t\ket t$ on $J$ qubits, with Qiskit's little-endian convention: qubit $q$ carries bit $b_q$ of the index $i=\sum_qb_q2^q$. Hence $i\in W_j$ if and only if $b_{J-1}=\dots=b_{j+1}=0$ and $b_j=1$: the level of an index is the position of its leading $1$.

\paragraph{Level shares.}
For a wavelet transform $\mathbf W$ and a vector $y$, let $E_j(y)=\|P_j\mathbf Wy\|^2$, with $P_j$ the projector onto $W_j$, and let the level share be $p_j(y)=E_j(y)/\|y\|^2$. A computational-basis measurement of $\mathbf W\ket y$ lands in $W_j$ with probability $p_j(y)$.

\paragraph{Access model.}
A unitary $\A_B$ prepares $\ket B$ on $J-1$ qubits. Cost is measured in queries to the state-preparation oracle. Under \emph{sampling access}, each query applies $\A_B$ and is followed by a measurement. Under \emph{coherent access}, a query is one application of $\A_B$ or of $\A_B^\dagger$, as amplitude estimation requires; controlled versions are not assumed. Every query prepares the same record, so all measurements interrogate one realization of the process. Such an oracle arises when the path is the output of a quantum simulation of the process, as in the spectral construction of fBm paths by \citet{BoulandDandapaniPrakash2023}, or of a quantum sensor; for classical data, loading the record is itself a state-preparation problem \citep{PleschBrukner2011,ZhangLiYuan2022}.

\section{Circuits}\label{sec:circuits}

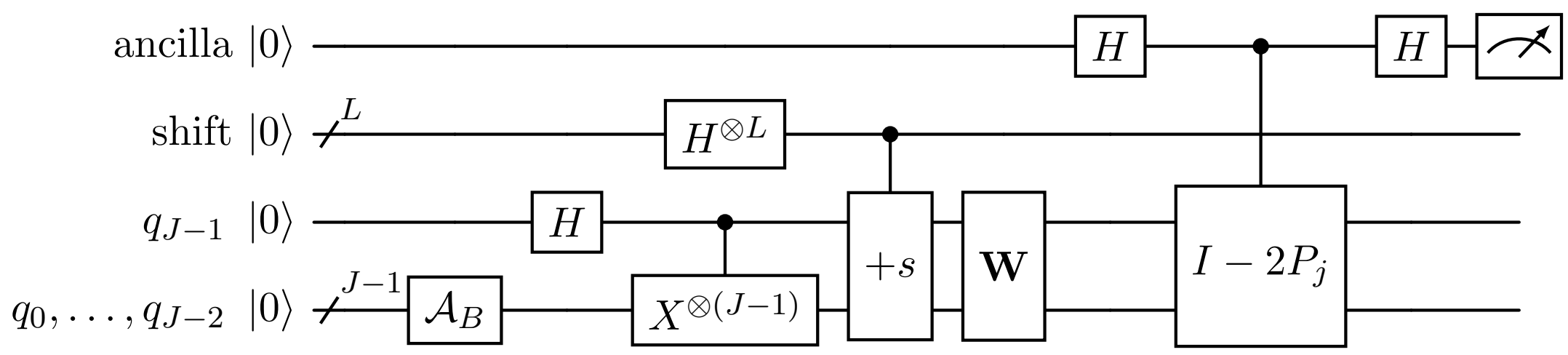
\begin{figure}[t]
\centering
\begin{quantikz}[row sep={0.8cm,between origins},column sep=0.28cm]
\lstick{ancilla $\ket0$} & & [3mm]& & & & & \gate{H} & \ctrl{2} & \gate{H} & \meter{}\\
\lstick{shift $\ket0$} & \qwbundle{L} & & & \gate{H^{\otimes L}} & \ctrl{1} & & & & &\\
\lstick{$q_{J-1}\ \ket0$} & & & \gate{H} & \ctrl{1} & \gate[2]{+s} & \gate[2]{\mathbf W} & & \gate[2]{I-2P_j} & &\\
\lstick{$q_{0},\dots,q_{J-2}\ \ket0$} & \qwbundle{J-1} & \gate{\A_B} & & \gate{X^{\otimes(J-1)}} & & & & & &
\end{quantikz}
\caption{The level test. $\A_B$ loads the path; the Hadamard gate on $q_{J-1}$ and the controlled $X$ gates form the mirror extension; the optional shift register adds a uniform shift $s$ for the nondecimated transform; $\mathbf W$ is the lattice wavelet transform; and the ancilla reads the reflection about level $j$. The probability of ancilla outcome $1$ is the level share $p_j$.}
\label{fig:pipeline}
\end{figure}

The circuit prepares $\ket B$, mirrors it, applies an optional shift, and performs the wavelet transform; Figure~\ref{fig:pipeline} shows the stages together with the level test of Section~\ref{sec:hadamard}. Only the stage $\mathbf W$ depends on the filter, and only the final reflection depends on the level being read.

\subsection{Mirror Extension}\label{sec:mirror}

\begin{proposition}[Mirror extension]\label{prop:mirror}
Adjoin qubit $J-1$, apply a Hadamard gate to it, and apply $X$ to each of qubits $0,\dots,J-2$ controlled on qubit $J-1$. This maps $\ket0\ket B$ to the normalized state of
$$
\tilde B=(B_0,B_1,\dots,B_{n-1},B_{n-1},\dots,B_1,B_0)\in V_J .
$$
The circular differences of $\tilde B$ are $(x_1,\dots,x_{n-1},0,-x_{n-1},\dots,-x_1,0)$, so the periodic extension of $\tilde B$ has no jump. Consequently, for any filter $g$ of length $\ell$ with $\sum_rg_r=0$, every circular translate satisfies
$$
\E\langle g_{\tau},\tilde B\rangle^2\le\sigma^2\|g\|_1^2\,(\ell-1)^{2H}\le\sigma^2\,\ell\,(\ell-1)^{2H}\|g\|_2^2,
$$
uniformly in $\tau$ and in $N$. Without the mirror, a translate that wraps around the end of $B$ has second moment of order $\sigma^2(\sum_{r\ge r_0}g_r)^2n^{2H}$, where $r_0$ is the wrap point.
\end{proposition}

\begin{proof}
The controlled $X$ gates map $t\mapsto n-1-t$ on the branch where qubit $J-1$ is $1$, which gives the stated vector with norm $\sqrt2\|B\|$. Its circular differences follow by inspection. Because $\sum_rg_r=0$, $\langle g_\tau,\tilde B\rangle=\sum_rg_r(\tilde B_{\tau+r}-\tilde B_\tau)$. Each difference $\tilde B_{\tau+r}-\tilde B_\tau$ equals $B_a-B_b$ for indices with $|a-b|\le r\le\ell-1$, because folding does not increase distances; its variance is at most $\sigma^2(\ell-1)^{2H}$. The Cauchy--Schwarz inequality gives the first bound, and $\|g\|_1^2\le\ell\|g\|_2^2$ the second. Without the mirror, a wrapping translate contains the difference $B_{n-1}-B_0$, of variance $\sigma^2(n-1)^{2H}$, multiplied by the tail sum of $g$.
\end{proof}

\begin{figure}[t]
\centering
\includegraphics[width=\linewidth]{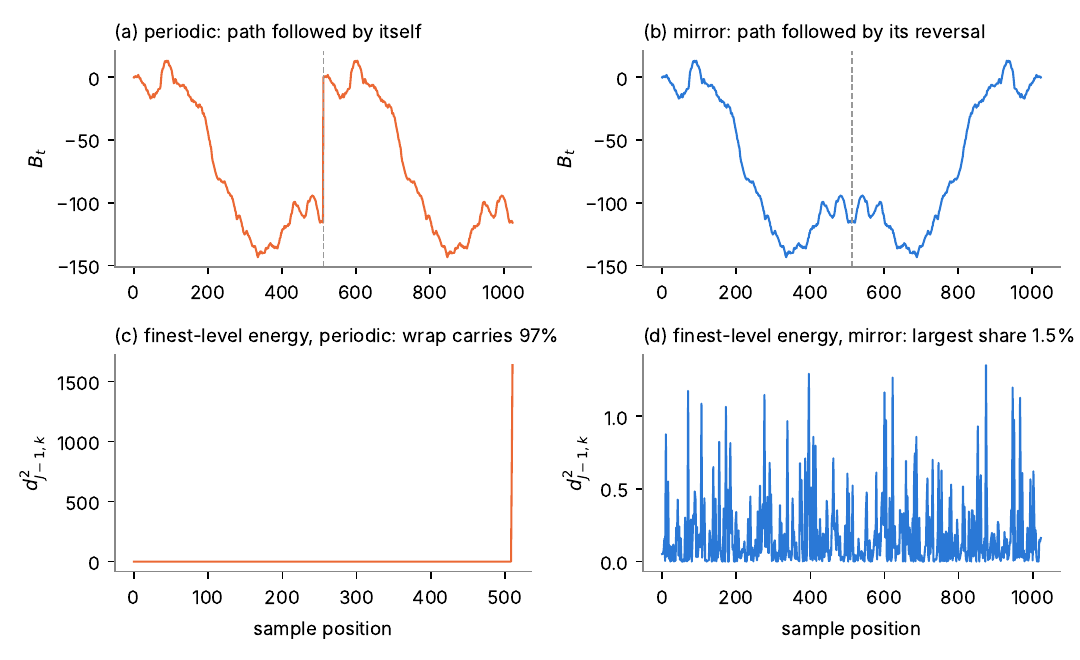}
\caption{A path with $H=0.8$ and $n=512$. Top: the periodic extension seen by circular filtering (a) has a jump at the seam, while the mirror extension (b) is continuous. Bottom: squared finest-level db2 coefficients. Under periodic handling (c) the coefficient straddling the seam carries almost all of the level energy; under the mirror (d) the energy is spread over the level.}
\label{fig:boundary}
\end{figure}

For a level-$j$ wavelet filter, $\ell=(2p-1)(2^{J-j}-1)+1$ and the interior coefficients have variance of order $\sigma^2\,2^{(2H+1)(J-j)}$, so the bound says that coefficients near a fold are no larger in order than interior ones. At most $2p-1$ decimated coefficients per level straddle each fold. For a fixed level offset, their expected energy as a fraction of expected level energy is $O(p\,2^{-j})$, with constants depending on the filter and offset. With periodic handling, wrapping coefficients with nonzero filter tail sums have second moments of order $n^{2H}$ at each fixed offset, and their aggregate share of the level energy tends to $1$ when $H>\tfrac12$. Figure~\ref{fig:boundary} shows a single path at $H=0.8$: under periodic handling one coefficient carries almost all of the finest-level energy, whereas under the mirror no coefficient carries more than a few percent.

\subsection{Wavelet Transform for an Arbitrary Filter}\label{sec:lattice}

Let $h$ be an orthonormal lowpass filter of length $2p$ and $g_k=(-1)^kh_{2p-1-k}$ the highpass filter. One analysis step on a periodic sequence of even length is $a_k=\sum_th_tu_{2k+t}$, $d_k=\sum_tg_tu_{2k+t}$. In polyphase form, with even and odd samples $(u_{2k},u_{2k+1})$,
$$
\binom{a_k}{d_k}=\sum_{m=0}^{p-1}E_m\binom{u_{2(k+m)}}{u_{2(k+m)+1}},\qquad E_m=\begin{pmatrix}h_{2m}&h_{2m+1}\\ g_{2m}&g_{2m+1}\end{pmatrix}.
$$
The polyphase matrix $E(z)=\sum_mE_mz^m$ is paraunitary, and it factors as
\begin{eqnarray}\label{eq:lattice}
E(z)=R_{p-1}\Lambda(z)R_{p-2}\Lambda(z)\cdots\Lambda(z)R_0,\qquad\Lambda(z)=\operatorname{diag}(1,z),
\end{eqnarray}
with $R_m$ real orthogonal $2\times2$ matrices \citep{VaidyanathanHoang1988,Vaidyanathan1993}. The factors are found by peeling: $R_{p-1}$ is the orthogonal matrix whose first column annihilates the column space of $E_{p-1}$, and $\Lambda(z)^{-1}R_{p-1}^\top E(z)$ is a paraunitary matrix of degree $p-2$. For db2 the lattice is a reflection at angle $-120^\circ$ followed by a rotation by $165^\circ$; for db4 it has four factors. A rotation mixes only the two samples of a pair; the delay $\Lambda(z)$ moves the odd samples by one pair, so that the next rotation combines samples from neighbouring pairs and the filter support grows by two with each stage.

\begin{proposition}[Lattice circuit]\label{prop:lattice}
Producing $W_j$ from the approximation block of $V_{j+1}$, that is, from the indices whose qubits $j+1,\dots,J-1$ are $\ket0$, proceeds as follows. Apply $R_0$ to qubit $0$. For $m=1,\dots,p-1$, apply $\mathrm{DEC}_{1..j}$ controlled on qubit $0$ being $\ket1$, followed by $R_m$ on qubit $0$. Finally, apply $\mathrm{ROT}_{0..j}$. All operations carry open controls on qubits $j+1,\dots,J-1$. Here $\mathrm{DEC}_{1..j}$ decrements the position register modulo $2^j$, and $\mathrm{ROT}_{0..j}$ cyclically rotates qubits $0,\dots,j$ to move qubit $0$ to position $j$. Repeating the procedure for $j=J-1,J-2,\dots,J-L$ implements the periodic orthonormal wavelet transform with $L$ levels in Mallat order. It uses $pL$ one-qubit rotations (with the specified controls), $(p-1)\sum_jj$ multiply-controlled $X$ gates, and $\sum_jj$ swaps with the specified controls, where $\sum_{j=J-L}^{J-1}j=L(2J-L-1)/2$.
\end{proposition}

\begin{proof}
On the active block, qubit $0$ holds the polyphase index and qubits $1,\dots,j$ hold $k$. A single-qubit gate on qubit $0$ applies a $2\times2$ matrix to each pair $(u_{2k},u_{2k+1})$. The decrement controlled on qubit $0$ moves the amplitude of $\ket{k+1,1}$ to $\ket{k,1}$, which is the advance $\Lambda(z)$ of the odd channel. By \eqref{eq:lattice} the sequence produces $a_k$ on $\ket{k,0}$ and $d_k$ on $\ket{k,1}$. The rotation moves the channel bit to qubit $j$, placing $d_k$ at index $2^j+k$ and $a_k$ at index $k$, which is Mallat order. Open controls leave the finer levels untouched. The counts follow from the construction.
\end{proof}

\begin{figure}[t]
\centering
\begin{subfigure}[b]{0.48\linewidth}\centering
\begin{quantikz}[row sep={0.72cm,between origins},column sep=0.36cm]
\lstick{$q_3$} & & \gate[3]{\mathrm{DEC}} & & & & \swap{1} & \\
\lstick{$q_2$} & & & & & \swap{1} & \targX{} & \\
\lstick{$q_1$} & & & & \swap{1} & \targX{} & & \\
\lstick{$q_0$} & \gate{R_0} & \ctrl{-1} & \gate{R_1} & \targX{} & & &
\end{quantikz}
\caption{finest level $j=3$ of db2}
\end{subfigure}\hfill
\begin{subfigure}[b]{0.48\linewidth}\centering
\begin{quantikz}[row sep={0.72cm,between origins},column sep=0.42cm]
\lstick{$q_3$} & \targ{} & & & \\
\lstick{$q_2$} & \octrl{-1} & \targ{} & & \\
\lstick{$q_1$} & \octrl{-1} & \octrl{-1} & \targ{} & \\
\lstick{$q_0$} & \ctrl{-1} & \ctrl{-1} & \ctrl{-1} &
\end{quantikz}
\caption{controlled decrement of $q_1,q_2,q_3$}
\end{subfigure}
\caption{(a) One level of the lattice circuit for db2 on $J=4$ qubits: the rotation $R_0$ on the channel qubit, the delay as a decrement of the position register controlled on $q_0=\ket1$, the rotation $R_1$, and the cyclic rotation that moves the channel bit to $q_3$. Coarser levels repeat the pattern on fewer qubits, with open controls on the qubits above. (b) The decrement: a bit flips when all lower bits of the register are $0$, processed from the top bit down.}
\label{fig:lattice}
\end{figure}
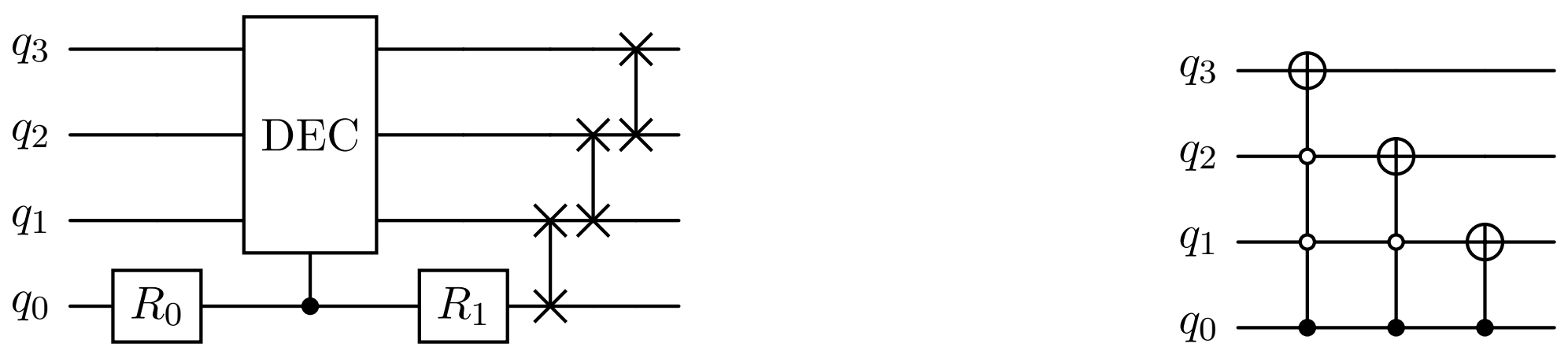

Figure~\ref{fig:lattice} shows the finest level of a db2 transform on four qubits and the decomposition of its decrement into multiply-controlled $X$ gates. A decrement on $j$ qubits is $j$ such gates, so the transform needs $O(pLJ)$ gates with at most $J-1$ controls each, at this level of abstraction. Elementary gate counts depend on how the controlled gates are decomposed and on the ancilla and connectivity assumptions; Section~\ref{sec:resources} reports counts under explicit assumptions. For the Haar filter, $p=1$ and the construction reduces to the familiar Hadamard-and-rotate circuit. \citet{BagherimehrabAspuruGuzik2024} implement wavelet transforms of any order differently, as a linear combination of unitaries, most of them permutations, which succeeds with a known probability and is made deterministic by amplitude amplification; the lattice circuit is deterministic by construction and uses no ancilla beyond those of the multiply-controlled gates. The construction was verified against the filter-bank matrix for db1--db4 at $J=7$, $L=3$, with maximum error $8\times10^{-16}$.

\subsection{Nondecimated Transform by a Shift Register}\label{sec:shift}

The nondecimated transform is the $\epsilon$-decimated QNDWT of \citet{Vidakovic2025QNDWT}, which we reuse with two changes: the shifted record is the mirrored path, and the transform is the lattice circuit of Section~\ref{sec:lattice}. We restate the construction in the form needed for the readouts below.

\begin{proposition}[Shift register \citep{Vidakovic2025QNDWT}]\label{prop:shift}
Adjoin an $L$-qubit register, prepare it in $H^{\otimes L}\ket0$, and add its value to the data register modulo $N$ (for bit $b$ of the shift register, an increment of qubits $b,\dots,J-1$ controlled on that bit). Then apply the wavelet transform to the data register. The reduced state of the data register is $2^{-L}\sum_s\mathbf WS^s\ket{\tilde B}\bra{\tilde B}S^{-s}\mathbf W^\dagger$, and its level shares are the nondecimated shares
$$
\bar p_j=2^{-L}\sum_{s=0}^{2^L-1}p_j(S^s\tilde B)=\frac{2^{j-J}\sum_{\tau\in\mathbb Z_N}\langle g_{j,\tau},\tilde B\rangle^2}{\|\tilde B\|^2},
$$
where $g_{j,\tau}$ is the level-$j$ equivalent filter starting at $\tau$.
\end{proposition}

\begin{proof}
The register is never touched again, so tracing it out averages over $s$. The second equality holds because, as $k$ and $s$ vary, the start $2^{J-j}k-s$ of the level-$j$ filter hits every residue modulo $N$ equally often.
\end{proof}

What the mirror adds to this construction is that every circular shift of $\tilde B$ is as harmless as the identity: Proposition~\ref{prop:mirror} holds for all $N$ translates, so the shift-averaged energies of a path carry no endpoint jump and no trimming is needed. The shift register adds $L$ qubits and $O(LJ)$ controlled gates, and a single transform acts coherently across the shift branches. The construction was verified to $3\times10^{-16}$.

\section{Hadamard Tests of Level Reflections}\label{sec:hadamard}

The Hadamard test estimates $\operatorname{Re}\bra\psi U\ket\psi$ with one ancilla \citep{Aharonov2009}: a Hadamard gate on the ancilla, $U$ controlled on it, a second Hadamard gate, and a measurement of the ancilla, so that $\Pr(\text{ancilla}=1)=\tfrac12(1-\operatorname{Re}\bra\psi U\ket\psi)$. Inserting $S^\dagger$ before the second Hadamard gate gives the imaginary part.

\begin{lemma}[Only the reflection is controlled]\label{lem:ht}
Let $\ket\psi=\A\ket0$ with $\A$ the preparation, mirror, optional shift and wavelet stages, and let $D$ be diagonal in the computational basis. The Hadamard test of $U=\mathbf W^\dagger D\mathbf W$ on $\mathbf W^\dagger\ket\psi$ has the same ancilla statistics as the Hadamard test of $D$ on $\ket\psi$. Hence the wavelet circuit is applied once, uncontrolled, and is never inverted.
\end{lemma}

\begin{proof}
The controlled version of $\mathbf W^\dagger D\mathbf W$ is $(I\otimes\mathbf W^\dagger)\,(\text{controlled-}D)\,(I\otimes\mathbf W)$, because $\mathbf W^\dagger\mathbf W=I$ on the uncontrolled branch. The final $I\otimes\mathbf W^\dagger$ acts on the data register only and does not change the distribution of the ancilla.
\end{proof}

\begin{proposition}[Three tests]\label{prop:tests}
Let $\rho$ be the wavelet-domain state, either $\mathbf W\ket y\bra y\mathbf W^\dagger$ or the reduced data state in Proposition~\ref{prop:shift}, and write $p_j=\tr(P_j\rho)$.
For the level test, $D=I-2P_j$, the ancilla outcome $1$ has probability $p_j$. The controlled reflection applies $Z$ to qubit $j$ with a closed control on the ancilla and open controls on qubits $j+1,\dots,J-1$. Conditional on outcome $1$, the data state is $P_j\rho P_j/p_j$; for a pure input, this retains coherence across positions within the selected level.
For the threshold test, let $\Pi_c$ project onto states with the $c$ most significant qubits all in $\ket0$, where $1\le c\le L$. For $D=I-2\Pi_c$, the ancilla outcome $1$ has probability $1-\sum_{j>J-1-c}p_j$, which includes the share in $V_{J-L}$. The controlled reflection applies $Z$ to the ancilla with $c$ open controls.
For the phase test, $D=\sum_je^{i\omega j}P_j+(I-\sum_jP_j)$, the two quadratures recover $1+\sum_j(e^{i\omega j}-1)p_j$. This is the characteristic function obtained by assigning index $j$ to detail level $W_j$ and index zero to the smooth space. The controlled operation uses one phase gate per level, with the controls of the level test.
\end{proposition}

\begin{proof}
For a reflection $D=I-2P$, $\bra\psi D\ket\psi=1-2\|P\ket\psi\|^2$, so $\Pr(1)=\|P\ket\psi\|^2$. The gate descriptions follow from the leading-one characterization of $W_j$ in Section~\ref{sec:setting}. For the phase test, the orthogonal level decomposition gives $\tr(D\rho)=\sum_je^{i\omega j}p_j+1-\sum_jp_j$. The post-measurement state is that of a projective measurement of $\{P_j,I-P_j\}$.
\end{proof}

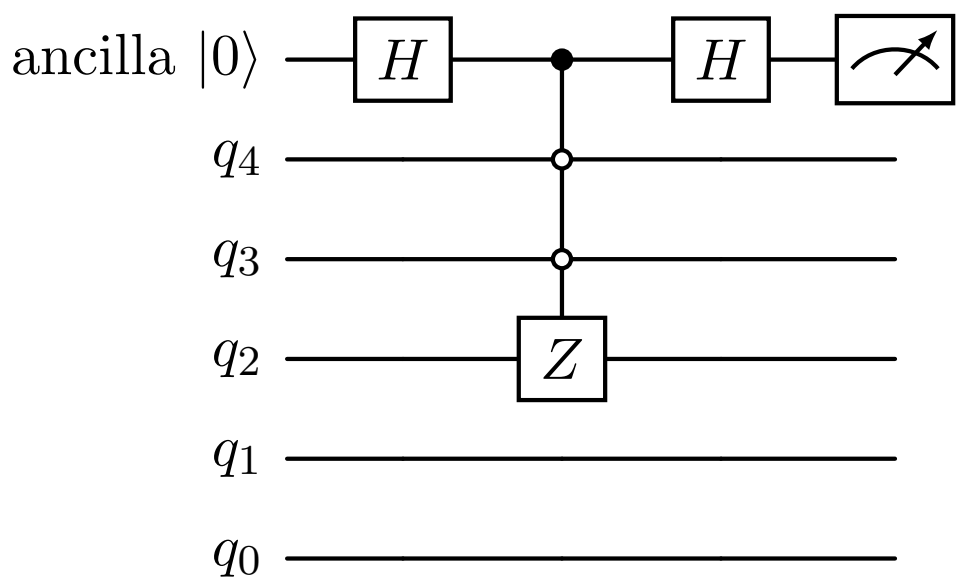
\begin{figure}[t]
\centering
\begin{quantikz}[row sep={0.66cm,between origins},column sep=0.45cm]
\lstick{ancilla $\ket0$} & \gate{H} & \ctrl{1} & \gate{H} & \meter{}\\
\lstick{$q_{4}$} & & \octrl{1} & & \\
\lstick{$q_{3}$} & & \octrl{1} & & \\
\lstick{$q_{2}$} & & \gate{Z} & & \\
\lstick{$q_{1}$} & & & & \\
\lstick{$q_{0}$} & & & &
\end{quantikz}
\caption{Level test for $W_2$ on $J=5$ data qubits after the wavelet transform. The $Z$ gate acts on $q_2$ when the ancilla is $\ket1$ and $q_4=q_3=\ket0$, which is the reflection $I-2P_2$. The same gate serves every filter.}
\label{fig:reflection}
\end{figure}

All three tests are independent of the wavelet; only the stage $\mathbf W$ changes. The use of level projectors to recover level energies goes back to \citet{Vidakovic2025QNDWT}; the level test makes it a specific readout of the $\epsilon$-decimated QNDWT, a reflection about an entire wavelet level after the transform, whose ancilla success probability is exactly the level share. They were verified against the classical shares and the characteristic function for db2 with mirror extension to $5\times10^{-16}$. Figure~\ref{fig:reflection} shows the controlled reflection for one level; the open controls select the indices whose leading $1$ is in bit $j$.

\begin{remark}[Phase operators in the time basis]\label{rem:phase}
The level test differs from the time-domain diagonal-phase probe described by \citet{Vidakovic2025QNDWT}, which applies $\operatorname{diag}(e^{i\theta\psi_{j,k}[t]})$ for a single wavelet atom directly to $\ket y$. The real-quadrature Hadamard test of that operator gives $1-\operatorname{Re}\bra yU\ket y\approx\tfrac{\theta^2}2\sum_t|\langle t\mid y\rangle|^2\psi_{j,k}[t]^2$, a $\psi^2$-weighted local energy of the record, whereas the normalized coefficient energy is $|\langle\psi_{j,k},y\rangle|^2$. The two differ already for a constant vector, whose coefficients vanish. The coefficient energy of an atom, and the energy share of a level, are obtained from reflections about basis indices after $\mathbf W$, as in Proposition~\ref{prop:tests}.
\end{remark}

\subsection{Readouts and Resources}\label{sec:resources}

A level test is a projective measurement of $\{P_j,I-P_j\}$ read out through one ancilla. Its statistics coincide with those of measuring the top $J-j$ qubits and checking the leading-one pattern, so with ideal readout it carries the same information about $p_j$. A direct measurement of all data qubits, however, returns the level of every shot, and therefore informs all levels of the window at once, whereas a level test spends each preparation on one specified level. The advantages of the test are operational: a single qubit is read, the data register survives in the measured level, and the controlled reflection is exactly the oracle needed for amplitude estimation (Section~\ref{sec:cost}), in which the data register must not be measured at all. Section~\ref{sec:readout} compares the three readouts at equal preparation cost.

Table~\ref{tab:resources} gives CNOT counts for the stages of Figure~\ref{fig:pipeline} at $J=10$ and $L=6$. No transpiler was used. Each multiply-controlled $X$ or $Z$ gate with $c\ge3$ controls is decomposed into $2c-3$ Toffoli gates with $c-2$ clean ancillas, and each Toffoli gate into six CNOT gates; a gate with two controls costs six CNOT gates and one with one control costs one. A multiply-controlled single-qubit rotation computes the conjunction of its controls into an ancilla, applies a singly controlled rotation (two CNOT gates), and uncomputes. Open controls add only single-qubit gates. Because nearly all gates of the transform share the channel qubit, the CNOT depth of the transform equals its count under this decomposition. The record preparation $\A_B$ is not included; a generic loading of $2^{9}$ amplitudes needs several hundred CNOT gates \citep{PleschBrukner2011}, its depth can be reduced at the price of ancillas \citep{ZhangLiYuan2022}, and it is the same in every protocol.

Three comparisons follow from the table. First, the controlled reflection of a level test costs between $1$ and $54$ CNOT gates across the window, at most $1.6\%$ of the db2 transform alone, so the comparison between a level test and direct readout turns on the number of qubits that must be read reliably and on the number of preparations, not on circuit size. Second, amplitude estimation is costed per query, a query being one application of $\A$ or of $\A^\dagger$; this is the unit of Table~\ref{tab:readout}. We use maximum-likelihood amplitude estimation \citep{Suzuki2020}, which applies powers of the Grover iterate $Q_j=-\A S_0\A^\dagger(I-2P_j)$ without control, so $\A$ itself is never controlled. The zero reflection $S_0$ must act on every register that $\A$ prepares: the ten data qubits for the decimated transform, and those together with the six shift qubits for the nondecimated transform used in Section~\ref{sec:readout}, which gives $15$ controls and $162$ CNOT gates. The uncontrolled level reflection costs $0$ to $42$ CNOT gates. Excluding $\A_B$, one Grover iterate of the nondecimated db2 circuit therefore costs $2\times5037+162$ plus at most $42$ CNOT gates, between $5118$ and $5139$ per query, against $5037$ for one preparation followed by measurement: per query, amplitude estimation is $1.6$--$2.0\%$ more expensive in CNOT gates, and both include the same $\A_B$. Parity per query is not parity in depth. A circuit at Grover power $m$ applies $\A$ followed by $Q_j^m$ coherently, that is, $2m+1$ queries in sequence, or $5037+m(10236\text{--}10278)$ CNOT gates excluding $\A_B$; with the largest powers used in Section~\ref{sec:readout}, $m=32$, $256$ and $2048$, this is about $3.3\times10^5$, $2.6\times10^6$ and $2.1\times10^7$ CNOT gates, against $5037$ for a circuit of direct readout. Third, canonical amplitude estimation with a phase register would require controlled powers of $Q_j$, hence a controlled $\A$ in which every gate acquires one more control; its gate cost per query is substantially larger, and we do not use it in the comparisons.

\begin{table}[t]
\centering\small
\begin{tabular}{l rr}
\toprule
stage ($J=10$, $L=6$) & CNOT gates & qubits read\\
\midrule
mirror extension & 9 & --\\
lattice transform, db1 & 1069 & --\\
lattice transform, db2 & 3396 & --\\
lattice transform, db4 & 8050 & --\\
shift register, $L=6$ & 1632 & --\\
controlled level reflection (level test), $j=9$ / $7$ / $4$ & 1 / 18 / 54 & 1 (ancilla)\\
controlled threshold reflection, $c=6$ & 54 & 1 (ancilla)\\
uncontrolled level reflection (Grover iterate), $j=9$ / $7$ / $4$ & 0 / 6 / 42 & --\\
zero reflection $S_0$, decimated (10 qubits) & 90 & --\\
zero reflection $S_0$, nondecimated (16 qubits) & 162 & --\\
Grover iterate, nondecimated db2, $j=9$ / $4$ & 10236 / 10278 & --\\
direct readout of the window & 0 & 6 (leading qubits)\\
\bottomrule
\end{tabular}
\caption{CNOT counts of the circuit stages under the decomposition stated in Section~\ref{sec:resources}, excluding the record preparation $\A_B$, which every protocol applies once per query. A level test adds one controlled reflection and one ancilla to the transform; direct readout adds no gates but reads the six leading data qubits; a Grover iterate contains two queries.}
\label{tab:resources}
\end{table}

\section{Estimating the Hurst Exponent}\label{sec:hurst}

The level tests deliver estimates $\hat p_j$ of the shares on a window $\W$ of fine levels. This section states how they depend on $H$, how $H$ is fitted, how the fit behaves asymptotically, and what it costs.

\subsection{Calibration and Asymptotic Distribution}

For the mirrored path, $E_j(\tilde B)=B^\top A_jB$ with the symmetric matrix $A_j=M^\top\mathbf W^\top P_j\mathbf WM$ (decimated), or its average over shifts (nondecimated), where $M$ is the $N\times n$ mirror matrix. With $\Sigma(H)$ the path covariance at unit increment variance,
$$
e_j(H)=\E E_j(\tilde B)=\sigma^2\tr\{A_j\Sigma(H)\},\qquad\Gamma_{jk}(H)=\Cov\{E_j,E_k\}=2\sigma^4\tr\{A_j\Sigma(H)A_k\Sigma(H)\},
$$
which are exact for every wavelet, $N$ and window. In the interior scaling regime, $e_{j-1}/e_j\to2^{2H}$ as $J-j$ grows while the filter support remains small relative to the record \citep{Flandrin1992,AbryVeitch1998}, and the exact calibration removes the deviation at the fine levels, which is where most of the information lies.

The next result justifies the Gaussian intervals of the estimator and the record-scale cost targets of Section~\ref{sec:cost}. It concerns a window at fixed offsets from the finest level, so that $K$ is fixed and $N\to\infty$, and the parameter space is a compact interval $\Theta\subset(0,1)$ with $H$ in its interior. Identification requires three conditions on the calibrated contrasts in the limit window: (I1) $D\{u(H')-u(H)\}\ne0$ for $H'\ne H$ in $\Theta$; (I2) $Dd\ne0$, with $d=\partial_Hu$; and (I3) the contrast covariance $D\Omega_HD^\top$ is nonsingular. For the configurations of Section~\ref{sec:sim} we verified them on the calibration grid: the least-squares slope of $u(H)$ across the window is strictly decreasing in $H$, and the smallest eigenvalue of $D\Gamma D^\top$, normalized by $e_je_k$, is at least $3.7\times10^{-3}$.

\begin{theorem}[Asymptotic normality of mirrored log-energy contrasts]\label{thm:clt}
Let the filter have $p\ge2$ vanishing moments, let $\W=\{J-K,\dots,J-1\}$ with $K$ fixed, use the mirrored decimated transform or the mirrored nondecimated transform with fixed $L$, and assume (I1)--(I3).

\emph{(a) Exact shares.} As $n\to\infty$,
$$
\sqrt N\,\big(\log E_j(\tilde B)-\log e_j(H)\big)_{j\in\W}\ \xrightarrow{d}\ N_K(0,\Omega_H),\qquad\Omega_H=\lim_{N\to\infty}N\Big(\frac{\Gamma_{jk}(H)}{e_j(H)e_k(H)}\Big)_{j,k\in\W},
$$
the coefficients that straddle a fold contribute $O_p(N^{-1})$ to each $\log E_j$, and the limit holds jointly with the normalized path norm, from which it is independent (Lemma~\ref{lem:indep}). With exact shares, $\sqrt N(\widehat H-H)\to N\{0,(d^\top P_{\Omega}d)^{-1}\}$.

\emph{(b) Finite shots.} Suppose that the level tests are independent given the record, that the budgets $M_j=M_j(N)$ are deterministic, and that $M_jp_j(\tilde B)\to\infty$ in probability for every $j\in\W$. Let $\hat s^2=(\hat d^\top P_{\hat V}\hat d)^{-1}$ be the plug-in variance of Section~\ref{sec:estimator}, with the record term evaluated at a consistent pilot and the shot term at the measured shares. Then $(\widehat H-H)/\hat s\to N(0,1)$. If moreover $M_jp_j(\tilde B)/N\to\kappa_j\in(0,\infty]$ in probability, where $\kappa_j$ may be a random variable measurable with respect to the limiting record, then $\sqrt N(\widehat H-H)$ converges to a variance mixture of normal distributions with mixing variance $\big(d^\top P_{\Omega'}d\big)^{-1}$, $\Omega'=\Omega_H+\operatorname{diag}(1/\kappa_j)$, all in natural-log units.
\end{theorem}

\begin{proof}
\emph{Fold terms.} Split $E_j=E_j^{\rm int}+E_j^{\rm fold}$, where $E_j^{\rm fold}$ collects the coefficients whose support meets a fold. There are at most $2(2p-1)$ such decimated coefficients per level, or a fraction $O(p2^{J-j}/N)$ of the translates in the nondecimated average, and by Proposition~\ref{prop:mirror} each has second moment of the order of an interior coefficient. Since $e_j$ is of order $N$ times the interior coefficient variance at a fixed offset, $E_j^{\rm fold}/e_j=O_p(N^{-1})=o_p(N^{-1/2})$.

\emph{Interior terms.} Index every interior coefficient by the position in the original path $B$ at which its support starts: a coefficient of the first half of $\tilde B$ at position $t$ is the level-$j$ filter applied to $(B_t,B_{t+1},\dots)$, and a coefficient of the second half whose support covers the mirrored positions $N-1-t'-\ell_j+1,\dots,N-1-t'$ is the time-reversed level-$j$ filter applied to $(B_{t'},B_{t'+1},\dots)$. By summation by parts, both are filters of the stationary increments $x$ with $p-1\ge1$ vanishing moments. Group positions into consecutive blocks of $2^{K}$ samples, the stride of the coarsest level in the window. For the decimated transform, each block contains a fixed number $2^{K-(J-j)}$ of starting positions of level $j$ in each half, at offsets that repeat from block to block, so the vector $Y_b$ of all centred squared coefficients whose support starts in block $b$, over the $K$ levels and both halves, is a fixed quadratic functional of a stationary Gaussian vector sequence indexed by $b$. For the nondecimated transform every position starts a translate and the same construction applies with $2^{K}$ translates per block. The cross-covariances of the underlying coefficients decay as $|b|^{2H-2p}$, and their squares are summable because $4H-4p<-1$ for $p\ge2$ and $H<1$. The multivariate central limit theorem for functionals of Hermite rank two of a stationary Gaussian vector sequence \citep{BreuerMajor1983,Arcones1994} therefore applies to $\sum_bY_b$. Dividing by $e_j\asymp N$ and applying the delta method to the logarithm gives the first display, with the limiting covariance equal to the limit of the exact normalized covariance because the fold terms are negligible. Under (I1)--(I3), the minimum-distance estimator in contrast space is consistent and its linearization gives the exact-share limit.

\emph{Finite shots.} Write $p_j=E_j/\|\tilde B\|^2=(e_j/N^{2H+1})\{1+Z_{N,j}/\sqrt N\}/S_N$, with $Z_N$ and $S_N$ as in Lemma~\ref{lem:indep}. Up to a factor $1+O_p(N^{-1/2})$, the random part of every share is the common factor $1/S_N$, so the shot variances $(1-p_j)/(M_jp_j)$, and hence the GLS weights, are asymptotically functions of $S_N$. Given the record, $\log\hat p_j-\log p_j$ are independent across $j$ and, because $M_jp_j\to\infty$, asymptotically normal with these variances by the Lindeberg theorem. The normalization $\log\|\tilde B\|^2$ is common to all levels and is removed by the contrasts. Linearizing, $\widehat H-H=w^\top(\xi^{\rm rec}+\xi^{\rm shot})\{1+o_p(1)\}$ with $w=(d^\top P_Vd)^{-1}P_Vd$, where $\xi^{\rm rec}$ is $Z_N/\sqrt N$ on the log scale and, given the record, $\xi^{\rm shot}$ is approximately $N(0,D_N)$ with $D_N=\operatorname{diag}\{(1-p_j)/(M_jp_j)\}$. Conditioning on the record, the characteristic function of $(\widehat H-H)/(w^\top Vw)^{1/2}$ at $t$ is $\exp\{it\,w^\top\xi^{\rm rec}/(w^\top Vw)^{1/2}-\tfrac12t^2w^\top D_Nw/(w^\top Vw)\}$ up to $o_p(1)$. By Lemma~\ref{lem:indep}, $Z_N$ converges jointly with $S_N$ to a limit $Z$ independent of $S$, while $w$ and $D_N$ are asymptotically functions of $S_N$; taking expectations, conditionally on $S$ the limit is the characteristic function of $N(0,1)$, because $w^\top Vw$ is the sum of the record variance $w^\top\Omega w/N$ and the shot variance $w^\top D_Nw$. Hence $(\widehat H-H)/(w^\top Vw)^{1/2}\to N(0,1)$, and the plug-in $\hat s^2$ is consistent for $w^\top Vw$ relative to its own size, which gives the studentized limit. Under the additional condition, $N D_N\to\operatorname{diag}(1/\kappa_j)$ with $\kappa_j$ a function of $S$, and the same argument without studentization gives the variance mixture.
\end{proof}

The step in the proof that requires care is the joint behaviour of the energy fluctuations and the path norm, because the norm sets the random shot variance and the GLS weights. The next lemma settles it.

\begin{lemma}[Asymptotic independence of fine-level fluctuations and the path norm]\label{lem:indep}
Realize all records on one probability space by self-similarity: with $B_H$ a standard fBm on $[0,1]$, let $B^{(n)}_t=\sigma n^HB_H((t+1)/n)$, $t=0,\dots,n-1$, which has the law of the record for every $n$. Let $Z_N=\sqrt N\,(E_j(\tilde B)/e_j-1)_{j\in\W}$ and $S_N=N^{-(2H+1)}\|\tilde B\|^2$. Under the conditions of Theorem~\ref{thm:clt}, $(Z_N,S_N)\to(Z,S)$ in distribution, where $Z\sim N_K(0,\Omega_H)$, $S=2^{-2H}\sigma^2\int_0^1B_H(u)^2\,du$, and $Z$ and $S$ are independent. The same holds jointly with any finite vector of path values $(B_H(t_1),\dots,B_H(t_m))$.
\end{lemma}

\begin{proof}
Let $G=(B_H(t_1),\dots,B_H(t_m))$. Its components belong to the first Wiener chaos of the isonormal Gaussian process generated by $B_H$. Each component of $Z_N$, fold coefficients included, is a centred quadratic form in the Gaussian vector $B^{(n)}$ and therefore belongs to the second Wiener chaos. By Theorem~\ref{thm:clt}(a) and the delta method, $Z_N\to N_K(0,\Omega_H)$. Elements of different Wiener chaoses are uncorrelated, so the covariance matrix of $(G,Z_N)$ is block diagonal and converges to $\operatorname{diag}(\Cov G,\Omega_H)$. For vectors of multiple Wiener integrals of fixed orders, componentwise convergence to normal laws together with convergence of the covariance matrix implies joint convergence to the Gaussian law with the limiting covariance \citep{PeccatiTudor2005}. Hence $(G,Z_N)\to(G,Z)$ with $G$ and $Z$ jointly Gaussian and uncorrelated, that is, independent.
For the norm, $\|\tilde B\|^2=2\|B^{(n)}\|^2$ and $N=2n$ give $S_N=2^{-2H}\sigma^2n^{-1}\sum_{t<n}B_H((t+1)/n)^2$, a Riemann sum that converges almost surely to $S$ by continuity of the paths. For fixed $m$, the Riemann sum $S^{(m)}=2^{-2H}\sigma^2m^{-1}\sum_{i=1}^mB_H(i/m)^2$ is a continuous function of $G$ with $t_i=i/m$, so $(Z_N,S^{(m)})\to(Z,S^{(m)})$ with $Z$ independent of $S^{(m)}$, and $S^{(m)}\to S$ almost surely as $m\to\infty$. Since $\lim_m\limsup_N\Pr(|S_N-S^{(m)}|>\delta)=0$ for every $\delta>0$, the approximation theorem for weak convergence \citep[Theorem~3.2]{Billingsley1999} gives $(Z_N,S_N)\to(Z,S)$ with $Z$ independent of $S$.
\end{proof}

Part (b) of Theorem~\ref{thm:clt} covers the budgets used in practice. A deterministic budget proportional to $N^{1+2H}$ makes $M_jp_j/N$ converge to a random limit, because $N^{2H}p_j$ carries the random path norm; the unstudentized error is then a normal mixture, but the studentized statistic is standard normal and the intervals remain valid. The condition $M_jp_j\to\infty$ excludes sparse counts, where some levels have only a few successes; that regime is handled by the count fit of Section~\ref{sec:estimator}, whose calibration we check by simulation in Section~\ref{sec:sparse}.

With a Haar filter applied to a path, $p=1$ and the squared covariances are not summable when $H>\tfrac34$; the limit is then non-Gaussian and the rate slower \citep{Bardet2002}. This is one reason to use the lattice construction with $p\ge2$. The theorem also clarifies the role of the shot budget: when $M_jp_j$ grows at least linearly in $N$, shot noise adds a term of the same order as the record noise, and when $M_jp_j/N\to\infty$ it disappears.

\subsection{Estimator}\label{sec:estimator}

Let $z_j=\log_2\hat p_j$ for $j\in\W$, where $\hat p_j=C_j/M_j$ is the fraction of ancilla outcomes $1$ in $M_j$ independent level tests. The count model is binomial conditional on the path. Since $\log_2p_j=\log_2E_j-\log_2\|\tilde B\|^2$, the record normalization cancels in any log contrast. Let $D$ be a fixed full-row-rank contrast matrix with $D\one=0$ and $K-1$ rows, and let $u_j(H)=\log_2e_j(H)$. The delta method gives the first-order covariance of $Dz$ as $DVD^\top$, where
$$
V=\frac{1}{(\ln 2)^2}\left\{\left(\frac{\Gamma_{jk}}{e_je_k}\right)_{j,k\in\W}
+\operatorname{diag}_{j\in\W}\left(\frac{1-p_j}{M_jp_j}\right)\right\}.
$$
The trace term describes variation across records. The diagonal term is the conditional shot-noise contribution and depends on the realized shares. When every selected level has a positive count, we estimate $H$ by minimizing
$$
\{D[z-u(H)]\}^\top(DVD^\top)^{-1}\{D[z-u(H)]\}.
$$
This is equivalent to profiling out a free intercept $\alpha$ in a GLS fit of $z$ to $\alpha\one+u(H)$. Any covariance term of the form $\one a^\top+a\one^\top$ drops out in contrast space. In the two-step fit, $V$ is held fixed at its plug-in value, using an unweighted pilot for the record term and the measured shares for the shot term \citep{Rao1967}. For an interior estimate, the first-order variance is $(d^\top P_Vd)^{-1}$, where $d=\partial_Hu$ and $P_V=D^\top(DVD^\top)^{-1}D$. Estimates are constrained to $[0,1]$, with endpoint predictions interpreted by their limits where needed.

The log fit is undefined when a level has no recorded success, which is common at large $H$ because the fine-level shares are small. The count fit avoids the logarithm. Conditional on the record, the count $C_j$ of a level test is binomial with success probability $p_j$, and the window counts of direct readout are cells of a multinomial distribution; in both cases we start from independent Poisson counts with means $M_jp_j$, with $M_j$ the total number of queries for direct readout, and integrating the intercept below turns this into a conditional multinomial likelihood. Write $\log p_j=b_j(H)+\alpha+\eta_j$ with $b_j(H)=u_j(H)\ln2$. The common log-intercept $\alpha$ contains the normalization $\log\|\tilde B\|^2$ and is left unrestricted, with a flat prior. The vector $\eta$ lies in the contrast space $\one^\perp$ and carries the record deviations: $\eta=Zc$, with $Z$ a $K\times(K-1)$ orthonormal basis of $\one^\perp$ and $c\sim N(0,A_H)$, $A_H=Z^\top S_HZ$, where $S_H=(\Gamma_{jk}/(e_je_k))_{j,k\in\W}$ is the record covariance of the log energies. The intercept integrates in closed form. For $C_+=\sum_jC_j>0$,
$$
\int_{\mathbb R}\prod_{j\in\W}\frac{(M_je^{b_j+\alpha+\eta_j})^{C_j}}{C_j!}\,e^{-M_je^{b_j+\alpha+\eta_j}}\,d\alpha=\frac{\Gamma(C_+)}{\prod_jC_j!}\prod_{j\in\W}\pi_j^{C_j},\qquad\pi_j=\frac{M_je^{b_j(H)+\eta_j}}{\sum_kM_ke^{b_k(H)+\eta_k}},
$$
which is a multinomial likelihood conditional on $C_+$ up to a factor that does not depend on $H$. The integrated likelihood is therefore
$$
L(H)=\int_{\mathbb R^{K-1}}\prod_{j\in\W}\pi_j(H,Zc)^{C_j}\,\phi_{A_H}(c)\,dc,
$$
with $\phi_A$ the $N(0,A)$ density, and only the $K-1$ contrast dimensions require numerical integration. We use $L(H)$ as a working likelihood and separate its exact and approximate parts. Its readout factor is exact for direct readout: given the record and the number $C_+$ of shots that land in the window, their levels are multinomial with probabilities $p_j/\sum_{k\in\W}p_k$. For level tests it rests on the Poisson approximation to the binomial counts, which is accurate when the shares are small and slightly conservative when they are not. The record factor is approximate in two ways. First, the Gaussian law of $c$, with the exact covariance $A_H$, is a lognormal approximation to the level energies that Theorem~\ref{thm:clt}(a) justifies only asymptotically. Second, $L(H)$ mixes over the unconditional law of $c$, whereas the exact likelihood conditional on $C_+$ would mix over the law of $c$ given $C_+$. Under the flat prior on $\alpha$ the two coincide, but the actual normalization is not flat, and $C_+$ is larger for records with more energy in the window, which depends on $c$ through $\sum_ke^{b_k+\eta_k}$. The two mixtures therefore differ in finite samples, and the coverage results of Section~\ref{sec:sparse} are evidence for the working likelihood rather than a consequence of exactness. We evaluate $L(H)$ by the Laplace approximation about the mode $\hat c$,
\begin{eqnarray*}
\log L(H)&\approx&\sum_jC_j\log\hat\pi_j-\tfrac12\hat c^\top A_H^{-1}\hat c-\tfrac12\log\det A_H\\
&&{}-\tfrac12\log\det\big\{C_+Z^\top(\operatorname{diag}\hat\pi-\hat\pi\hat\pi^\top)Z+A_H^{-1}\big\},
\end{eqnarray*}
maximize over $H$ on a grid of $49$ points with quadratic interpolation, and take as interval the set of $H$ within $1.92$ of the maximum. A joint Laplace approximation over $(\alpha,c)$ differs from this one only by the Stirling approximation of $\Gamma(C_+)$, which does not depend on $H$, so the two give identical estimates and intervals. As a direct check of the Laplace approximation in the sparse regime, we also evaluated $L(H)$ by importance sampling with a multivariate $t$ proposal on five degrees of freedom centred at the Laplace mode, with $20000$ draws per grid value; Section~\ref{sec:sparse} reports the comparison.

This construction handles the dependence between the normalization and the energies of the same record to first order. The normalization never enters the likelihood, because it is common to all levels and integrated out, and the record model is placed only on contrasts, which are invariant to it; the remaining dependence between the path norm and the contrast deviations vanishes asymptotically by Lemma~\ref{lem:indep}. Placing a proper Gaussian prior on the common component instead would double-count the normalization; in a check at $H=0.9$ with sparse counts it changed the estimates by $0.016$ on average.

If every count is zero, the Poisson factor $e^{-M_je^{b_j+\alpha+\eta_j}}$ tends to $1$ as $\alpha\to-\infty$, so the integrand approaches a positive limit along the intercept direction and the integral over $\alpha$ diverges. Such a record yields no estimate; we report it as all-zero and count it as uninformative. Otherwise the count fit returns an estimate; when the counts are sparse the likelihood is flat and the interval is wide, and we call an estimate uninformative when its interval is wider than $0.4$. The cutoff $1.92$ is the asymptotic one; Section~\ref{sec:sparse} reports its coverage separately for records with at least one zero count, where the asymptotics are least reliable. Each replicate of that simulation regenerates both the path and its readout, so the coverage is a record-level check of the interval at the true parameter.

\subsection{Cost}\label{sec:cost}

A fine-level share is small. For a path the energy concentrates at coarse levels; for a fixed offset from the finest level, $p_j$ is of order $N^{-2H}$ in probability, with a record-dependent multiplier from the random path norm. The scaling approximation across levels gives the additional factor $2^{2H(J-1-j)}$ away from the finest-scale discretization effects. In the simulations below, the median of the smallest share in a window of six levels at $N=1024$ is $2.8\times10^{-2}$ at $H=0.2$, $1.1\times10^{-3}$ at $H=0.5$, $2.3\times10^{-5}$ at $H=0.8$ and $5.1\times10^{-6}$ at $H=0.9$.

\begin{proposition}[Cost of the level shares]\label{prop:cost}
For direct level tests, a relative root mean squared error of $\epsilon$ requires approximately $M_j=\epsilon^{-2}(1-p_j)/p_j$ queries. Under coherent access, amplitude estimation with the Grover iterate $Q_j=-\A S_0\A^\dagger(I-2P_j)$ uses $O(\epsilon^{-1}p_j^{-1/2})$ queries to $\A$ or $\A^\dagger$ for constant success probability. Here $S_0=I-2\ket0\bra0$ on all preparation registers and $P_j$ acts on the data register. Independent repetitions and a median increase confidence to $1-\delta$ at a multiplicative cost of order $\log(1/\delta)$ \citep{Brassard2002,Suzuki2020,Grinko2021}. Variants with shallower circuits trade some of the quadratic gain for depth \citep{GiurgicaTiron2022}, and the quantum-signal-processing approach of \citet{RallFuller2023} attains near-optimal constants; the scaling in $\epsilon$ and $p_j$ is the same. The coherent bound is a target query budget for a given success probability $p_j$ of the reflection. Selecting a budget without knowing $p_j$ requires a lower bound or a stopping rule.
At the $N^{-1/2}$ record-noise scale of Theorem~\ref{thm:clt}, these budgets are of order $N^{1+2H}$ and $N^{1/2+H}$, respectively, for typical records at a fixed fine level.
\end{proposition}

\begin{proof}
The first statement follows from the binomial variance. Amplitude estimation with $M$ queries has error at most $2\pi\sqrt{p(1-p)}/M+\pi^2/M^2$ with probability at least $8/\pi^2$ \citep{Brassard2002}; an error $\epsilon p$ requires $M\asymp\epsilon^{-1}p^{-1/2}$. The record-scale budgets follow by setting $\epsilon\asymp N^{-1/2}$.
\end{proof}

These are upper bounds for the stated protocols, counted in queries. Amplitude estimation needs the inverse of the whole preparation, including the state preparation of the record; the level test needs neither inverses nor controlled wavelet circuits. Direct readout of the leading qubits needs neither, and each of its preparations returns the level of the shot, so every level of the window receives all $M$ preparations, whereas a level test assigns its preparations to one level. With equal allocation over $K$ level tests, the shot variance of each $\log\hat p_j$ is therefore $K$ times that of direct readout at the same total budget; the loss in Fisher information about $H$ is smaller than a factor $K$, because the record term does not change. Equal allocation is not optimal in general, and we did not optimize it.

\section{Numerical Study}\label{sec:sim}

\subsection{Design}

Paths were generated by circulant embedding \citep{DaviesHarte1987}. The main experiment uses $n=512$ and six values of $H$ between $0.2$ and $0.9$, with $R=500$ paths per value. Each path was mirrored to $N=1024$ ($J=10$) and transformed with $L=6$ levels; the window is the six finest levels $\{4,\dots,9\}$. The periodic comparison transforms the unmirrored path ($J=9$, window $\{3,\dots,8\}$). The mirrored configurations use db2 with decimated and nondecimated transforms and db4 with the nondecimated transform. The periodic configurations use db2 with both transforms. Level tests were simulated from their exact Bernoulli law, with budgets of $Q=\infty$ (record noise only), $6\times2^{14}$ and $6\times2^{18}$ uses split equally across the six level tests. A record with a level test of zero successes counts as a failure of the log fit; root-mean-squared errors (RMSE) are over the remaining records. The calibration $e(H),\Gamma(H)$ was computed exactly by traces on a grid of $25$ values of $H$ and interpolated. The square root of the Cram\'er--Rao variance bound for $H$ from $n=512$ samples, with $\sigma^2$ unknown, is given for reference.

Three further experiments use the same machinery. The record-length experiment repeats the exact-share fit for $n\in\{256,512,1024,2048\}$ and $H\in\{0.5,0.8,0.9\}$ with $400$ paths per setting, the decimated db2 transform, and the same six-level window at fixed offsets from the finest level; it compares mirror extension, periodic handling, and a classical interior-only fit that applies the periodic transform to the unmirrored path and discards every coefficient whose support crosses the seam, with its own exact calibration. The readout experiment uses $300$ mirrored nondecimated db2 records per value of $H\in\{0.5,0.65,0.8,0.9\}$ and total budgets of $2^{17}$, $2^{20}$ and $2^{23}$ queries, a query being one application of $\A$ or $\A^\dagger$; every readout is applied to the same records, so differences between readouts are paired. Direct readout spends every query on one measurement of the data register, which yields the level of every shot. Level tests spend $\lfloor Q/6\rfloor$ queries on each level. Amplitude estimation spends $\lfloor Q/6\rfloor$ queries on each level with maximum-likelihood amplitude estimation \citep{Suzuki2020}, one representative of the amplitude-estimation methods without phase estimation \citep{Grinko2021,GiurgicaTiron2022}: Grover powers $m=0,1,2,4,\dots,m_{\max}$, each followed by $N_{\rm sh}$ measurements of the good-state indicator, with $m_{\max}$ the largest power for which $N_{\rm sh}\ge100$ fits the budget. This gives $(m_{\max},N_{\rm sh})=(32,164)$, $(256,169)$ and $(2048,170)$, and the queries actually used are $99.8\%$, $99.8\%$ and $99.7\%$ of the three budgets. The share is estimated by maximizing the exact binomial likelihood over the amplitude angle, and its log enters the log fit with the plug-in variance from the Fisher information of the schedule. A level whose estimated share is exactly zero, which happens only when no good outcome is observed at any power, makes the log fit fail for that record, and failures are reported separately. Both the log fit and the count fit are applied to direct readout and to level tests. Monte Carlo standard errors of RMSE ratios between readouts are computed by a paired bootstrap over records.

\subsection{Boundary Energy and Record Length}

Table~\ref{tab:boundary} compares the share of level energy carried by the coefficients that straddle a boundary with the share of coefficients that straddle it. With mirror extension the two shares are of comparable magnitude at the reported settings. With periodic handling at $H=0.8$, the coefficients that wrap at the finest level carry $98\%$ of the level energy for db2 and $99\%$ for db4.

\begin{table}[t]
\centering\small
\begin{tabular}{ll c ccc}
\toprule
&& coefficient & \multicolumn{3}{c}{energy share}\\
\cmidrule(lr){4-6}
filter, boundary & level & share & $H=0.3$ & $0.5$ & $0.8$\\
\midrule
db2, periodic & finest & 0.002 & 0.052 & 0.429 & 0.980\\
 & finest $-2$ & 0.016 & 0.148 & 0.526 & 0.962\\
 & finest $-5$ & 0.125 & 0.442 & 0.637 & 0.907\\
db2, mirror & finest & 0.004 & 0.004 & 0.004 & 0.007\\
 & finest $-2$ & 0.031 & 0.030 & 0.032 & 0.045\\
 & finest $-5$ & 0.250 & 0.236 & 0.243 & 0.300\\
db4, periodic & finest & 0.006 & 0.100 & 0.627 & 0.993\\
 & finest $-5$ & 0.375 & 0.805 & 0.877 & 0.975\\
db4, mirror & finest & 0.012 & 0.007 & 0.007 & 0.007\\
 & finest $-5$ & 0.750 & 0.731 & 0.738 & 0.761\\
\bottomrule
\end{tabular}
\caption{Share of the level energy carried by coefficients whose support straddles a boundary (the wrap for periodic handling, the two folds for mirror extension), and the share of such coefficients, decimated transform, $n=512$, 300 paths. The coefficient share is the reference value when boundary and interior coefficients contribute equally per coefficient.}
\label{tab:boundary}
\end{table}

\begin{figure}[t]
\centering
\includegraphics[width=\linewidth]{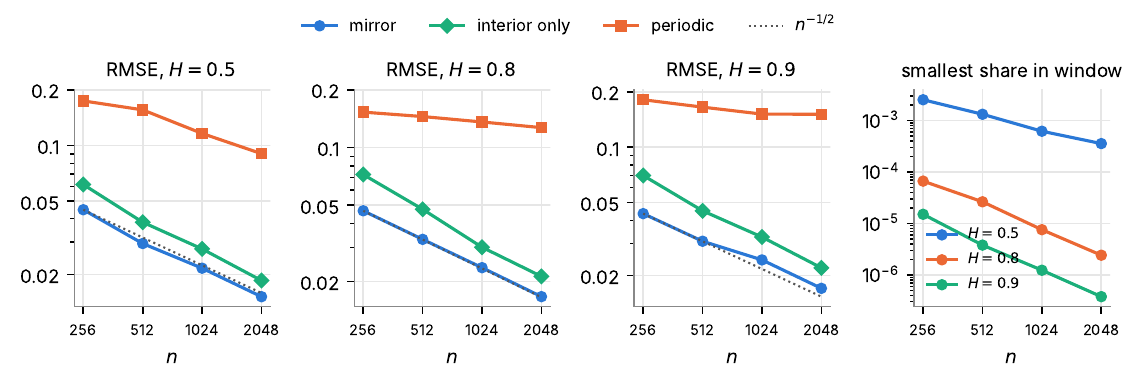}
\caption{Record-only RMSE of $\widehat H$ against record length $n$ for the mirrored transform, periodic handling, and the classical interior-only fit, db2 decimated, window of the six finest levels, $400$ paths per setting; the dotted line has slope $-1/2$. Right: median of the smallest level share in the window for the mirrored transform.}
\label{fig:lengths}
\end{figure}

Figure~\ref{fig:lengths} separates the effect of the jump from the effect of the boundary coefficients. With exact shares, the mirrored estimator has RMSE $0.045$, $0.029$, $0.022$ and $0.015$ at $H=0.5$ for $n=256$, $512$, $1024$ and $2048$, and $0.047$, $0.033$, $0.024$ and $0.017$ at $H=0.8$; fitted on the log scale, its RMSE decays like $n^{-0.51}$, $n^{-0.50}$ and $n^{-0.44}$ at $H=0.5$, $0.8$ and $0.9$, in agreement with the $N^{-1/2}$ rate of Theorem~\ref{thm:clt}. Under periodic handling the RMSE barely decreases with $n$ when $H\ge0.8$ (fitted exponents $-0.09$), stays above $0.12$ at $n=2048$, and is biased by $-0.036$ and $-0.078$ at $H=0.8$ and $0.9$: the jump does not average out, because its energy grows with $n$ as fast as the energy of the levels it contaminates.

The interior-only fit, which applies the periodic transform but discards the coefficients that cross the seam, removes the jump by other means. It also converges at about the $n^{-1/2}$ rate, but its RMSE is $23$--$61\%$ larger than that of the mirror at every length, because at the coarser levels of the window it discards a substantial fraction of the coefficients; at $n=256$ the coarsest level keeps only two of its four. The improvement of the mirror over periodic handling is therefore mostly the removal of the jump, and its further improvement over trimming comes from keeping every coefficient, with the folds handled by the exact calibration. Trimming also has no simple quantum counterpart, since discarding particular coefficients would require a reflection about a set of positions within each level.

The last panel of Figure~\ref{fig:lengths} shows the cost side. The median smallest share in the window decays like $n^{-0.96}$, $n^{-1.62}$ and $n^{-1.76}$ at $H=0.5$, $0.8$ and $0.9$, close to the $N^{-2H}$ rate of Section~\ref{sec:cost}. At $n=2048$ and $H=0.9$ it is $3.8\times10^{-7}$, so that a single level test succeeds about once in $2.6$ million preparations; this is the regime in which the quadratic reduction of amplitude estimation matters.

\subsection{Accuracy with Exact and Sampled Shares}

\begin{table}[t]
\centering\footnotesize\setlength{\tabcolsep}{4pt}
\begin{tabular}{l cccccc}
\toprule
RMSE of $\widehat H$ & $H=0.2$ & $0.35$ & $0.5$ & $0.65$ & $0.8$ & $0.9$\\
\midrule
\multicolumn{7}{l}{\emph{$Q=\infty$ (record noise only)}}\\
db2, mirror, decimated & 0.0237 & 0.0280 & 0.0330 & 0.0320 & 0.0325 & 0.0322\\
db2, mirror, nondecimated & 0.0230 & 0.0272 & 0.0313 & 0.0306 & 0.0303 & 0.0294\\
db4, mirror, nondecimated & 0.0238 & 0.0280 & 0.0329 & 0.0321 & 0.0322 & 0.0318\\
db2, periodic, decimated & 0.0273 & 0.0362 & 0.1404 & 0.1436 & 0.1426 & 0.1712\\
db2, periodic, nondecimated & 0.0248 & 0.0317 & 0.1222 & 0.0930 & 0.2011 & 0.2796\\
Root Cram\'er--Rao bound & 0.0214 & 0.0255 & 0.0275 & 0.0285 & 0.0286 & 0.0278\\
\midrule
\multicolumn{7}{l}{\emph{db2, mirror, nondecimated, level tests, log fit (failure rate)}}\\
$Q=6\times2^{18}$ & 0.0232 & 0.0282 & 0.0347 & 0.0451 & 0.0577 (11\%) & 0.0651 (40\%)\\
$Q=6\times2^{14}$ & 0.0247 & 0.0347 & 0.0512 (0.6\%) & 0.0725 (21\%) & 0.0907 (66\%) & 0.1096 (86\%)\\
\midrule
smallest share $p_j$ in window & $2.8\times10^{-2}$ & $6.8\times10^{-3}$ & $1.1\times10^{-3}$ & $1.8\times10^{-4}$ & $2.3\times10^{-5}$ & $5.1\times10^{-6}$\\
\bottomrule
\end{tabular}
\caption{Hurst estimation from level shares, $n=512$ ($N=1024$ after mirroring), window of six levels, $R=500$. Nominal 95\% intervals cover at $0.92$--$0.96$ for the mirrored configurations and at $0.64$--$0.97$ for periodic handling.}
\label{tab:hurst}
\end{table}

At $H\ge0.5$, the periodic configurations have much larger RMSE than the corresponding mirrored configurations (Table~\ref{tab:hurst}). The ratio is about $3$--$10$, depending on $H$ and the transform. Periodic estimates show biases as low as $-0.23$ and interval coverage as low as $0.64$. The boundary-energy results in Table~\ref{tab:boundary} show the concentration of energy that accompanies this loss of accuracy. With exact shares, the mirrored nondecimated db2 RMSE is approximately $6$--$14\%$ above the root Cram\'er--Rao variance bound computed from the full record, a reference for regular unbiased estimation. The mirrored db2 and db4 nondecimated configurations give comparable RMSEs. For db2, the nondecimated transform reduces RMSE by approximately $3$--$9\%$ relative to the decimated transform with exact shares, at the cost of the six qubits of the shift register.

Small fine-level shares limit the accuracy of the finite-shot estimates at large $H$. At $Q=6\times2^{14}$, a level test returns zero successes for $21\%$ of records at $H=0.65$ and $86\%$ at $H=0.9$. At $Q=6\times2^{18}$, the reported RMSE is close to the record-only RMSE through $H=0.5$, while the failure rates are still $11\%$ at $H=0.8$ and $40\%$ at $H=0.9$. Among records that yield estimates, the corresponding biases are $-0.013$ and $-0.033$, consistent with selection in favour of records with larger fine-level shares. The next two subsections address this regime from the two sides that are available: a readout that uses each preparation more efficiently, and a fit that does not discard records with zero counts.

\subsection{Readout at Equal Preparation Cost}\label{sec:readout}

\begin{figure}[t]
\centering
\includegraphics[width=\linewidth]{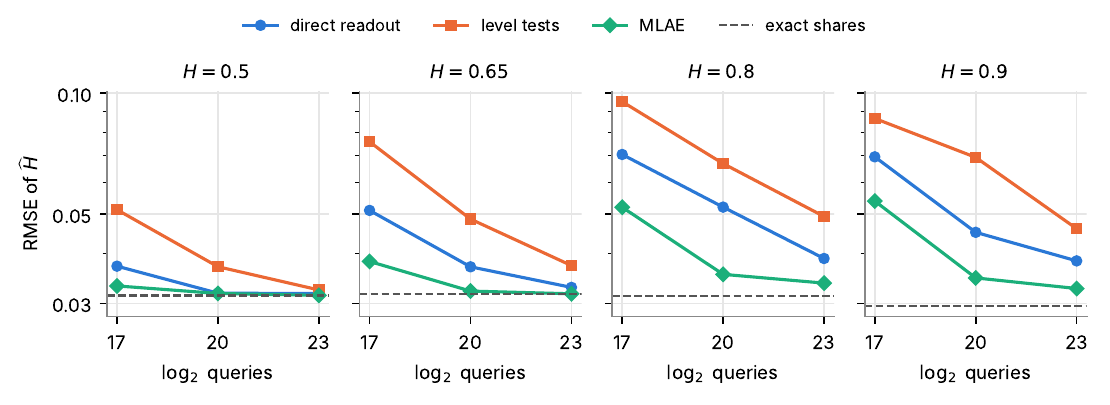}
\caption{RMSE against the number of queries to the preparation for direct readout and level tests (count fit) and for maximum-likelihood amplitude estimation (MLAE), with the record-only RMSE of the same records (dashed). Data of Table~\ref{tab:readout}.}
\label{fig:readout}
\end{figure}

Table~\ref{tab:readout} and Figure~\ref{fig:readout} compare the three readouts on the same records at equal numbers of queries, with the count fit for direct readout and level tests and the log fit for amplitude estimation; the column of exact shares is the record-only RMSE of the same records, and the ratios are paired. Direct readout is more accurate than separate level tests at every setting, with paired RMSE ratios between $0.65$ and $0.98$: at $2^{20}$ queries and $H=0.8$ its RMSE is $0.052$ against $0.067$. As stated in Section~\ref{sec:cost}, each preparation of direct readout returns the level of the shot and so serves all six levels, whereas a level test serves one; the shot variance of each level is six times larger for level tests, and the loss in accuracy is smaller than that because the record term is common to both. With ideal readout of the data register, a Hadamard level test is therefore not the efficient way to sample the shares.

Maximum-likelihood amplitude estimation, whose oracle is the level reflection, is at least as accurate as direct readout at every setting where it returns an estimate on essentially every record, and more accurate whenever the informative shares are small. At $H=0.9$ and $2^{20}$ queries its RMSE is $0.0347$, against $0.0451$ for direct readout and $0.0295$ with exact shares on the same records, a paired ratio to the record-only error of $1.18$ (SE $0.03$) and to direct readout of $0.77$ (SE $0.04$). With $2^{20}$ queries it is more accurate at $H=0.8$ and $0.9$ than direct readout with $2^{23}$. At $2^{23}$ queries it is within $11\%$ of the record-only error at every $H$, and at $H=0.5$ direct readout and amplitude estimation agree from $2^{20}$ queries on, while all three readouts agree to within $3\%$ at $2^{23}$. The one failure mode of amplitude estimation is a level for which no good outcome occurs at any Grover power; this happened on $16\%$ of records at $H=0.9$ with $2^{17}$ queries and on at most $0.3\%$ elsewhere. At that setting the RMSE of amplitude estimation and its paired advantage over direct readout refer to the $84\%$ of records on which it returns an estimate, and the comparison should be read with that restriction. The comparison is in queries. By Table~\ref{tab:resources}, a query of amplitude estimation costs $1.6$--$2.0\%$ more CNOT gates than a query of direct readout under the stated decomposition, so the two protocols have comparable ideal CNOT budgets per query. They do not have comparable depths: at $2^{20}$ queries the largest Grover power is $256$, and the longest circuit of amplitude estimation contains about $2.6\times10^6$ CNOT gates excluding $\A_B$, rising to $2.1\times10^7$ at $2^{23}$ queries with power $2048$, against $5037$ for a circuit of direct readout. The comparison therefore assumes ideal coherent queries; we did not repeat it at matched CNOT budgets or under noise, and a ranking at equal circuit cost or depth is not claimed.

\begin{table}[t]
\centering\footnotesize\setlength{\tabcolsep}{3.5pt}
\begin{tabular}{ll c cc c cc}
\toprule
&& exact & direct & level & & \multicolumn{2}{c}{RMSE ratio (SE)}\\
\cmidrule(lr){7-8}
$H$ & queries & shares & readout & tests & MLAE & MLAE/exact & MLAE/direct\\
\midrule
$0.5$ & $2^{17}$ & 0.0314 & 0.0371 & 0.0513 & 0.0332 & 1.06 (0.02) & 0.89 (0.03)\\
 & $2^{20}$ & 0.0314 & 0.0318 & 0.0370 & 0.0318 & 1.01 (0.01) & 1.00 (0.02)\\
 & $2^{23}$ & 0.0314 & 0.0318 & 0.0324 & 0.0314 & 1.00 (0.00) & 0.99 (0.01)\\
\addlinespace
$0.65$ & $2^{17}$ & 0.0317 & 0.0511 & 0.0757 [7\%] & 0.0382 & 1.20 (0.05) & 0.75 (0.04)\\
 & $2^{20}$ & 0.0317 & 0.0370 & 0.0486 & 0.0322 & 1.02 (0.02) & 0.87 (0.03)\\
 & $2^{23}$ & 0.0317 & 0.0329 & 0.0373 & 0.0317 & 1.00 (0.01) & 0.96 (0.01)\\
\addlinespace
$0.8$ & $2^{17}$ & 0.0314 & 0.0704 [0.7\%] & 0.0953 [23\%] & 0.0520 (0.3\%) & 1.67 (0.08) & 0.74 (0.03)\\
 & $2^{20}$ & 0.0314 & 0.0521 & 0.0668 & 0.0354 & 1.13 (0.03) & 0.68 (0.03)\\
 & $2^{23}$ & 0.0314 & 0.0388 & 0.0494 & 0.0337 & 1.07 (0.03) & 0.87 (0.04)\\
\addlinespace
$0.9$ & $2^{17}$ & 0.0295 & 0.0694 [3\%] & 0.0865 [26\%] & 0.0539 (16\%) & 1.80 (0.11) & 0.75 (0.04)\\
 & $2^{20}$ & 0.0295 & 0.0451 & 0.0692 [2\%] & 0.0347 & 1.18 (0.03) & 0.77 (0.04)\\
 & $2^{23}$ & 0.0295 & 0.0383 & 0.0460 & 0.0327 & 1.11 (0.03) & 0.85 (0.04)\\
\bottomrule
\end{tabular}
\caption{RMSE of $\widehat H$ for three readouts of the same mirrored nondecimated db2 records at equal numbers of queries to $\A$ or $\A^\dagger$, $n=512$, $300$ records per $H$. Direct readout and level tests use the count fit; brackets give the percentage of estimates whose interval is wider than $0.4$. MLAE is maximum-likelihood amplitude estimation with the log fit; parentheses give the percentage of records with a zero estimated share, excluded from its RMSE. ``Exact shares'' is the record-only RMSE of the same records. Ratios are computed on records where both fits succeed, with paired-bootstrap standard errors.}
\label{tab:readout}
\end{table}

\subsection{Sparse Counts}\label{sec:sparse}

The log fit discards every record on which a level has no success, and the records it keeps are those whose fine levels happen to carry more energy. Its failure rate is therefore informative in itself, and its conditional RMSE understates the error. Table~\ref{tab:sparse} lists the settings at which zero counts occur. Level tests with $2^{17}$ queries fail on $60\%$ of records at $H=0.8$ and $83\%$ at $H=0.9$, and the surviving estimates are biased by $-0.041$ and $-0.064$. Direct readout fails less often for the reason given above, but still on half of the records at $H=0.9$ with $2^{17}$ queries.

The count fit returns an estimate for every record in this experiment; no record had zero counts at every level, which is the one case without an estimate. Where the log fit fails on more than $10\%$ of records, the count fit reduces the bias by between a third and four fifths, its biases lie between $-0.004$ and $-0.027$ at every setting of the table, and its intervals cover at $0.93$--$0.97$ over all records. Among the records with at least one zero count, where the cutoff $1.92$ is least supported by asymptotics, coverage is $0.92$--$0.98$ in every setting with at least $40$ such records; the exception is level tests at $H=0.9$ with $2^{23}$ queries, where $21$ sparse records cover at $0.86$, within two binomial standard errors of $0.95$. When the counts are too sparse, the fit says so through the width of its interval: $23\%$ and $26\%$ of its intervals are wider than $0.4$ for level tests with $2^{17}$ queries at $H=0.8$ and $0.9$, against at most $3\%$ for direct readout. The Laplace approximation is not a source of error here: on $240$ records with at least one zero count, drawn from the three sparsest settings of Table~\ref{tab:sparse} for both readouts, the importance-sampling evaluation of the likelihood changes the estimate by $0.0017$ on average and by at most $0.023$, and moves the interval endpoints by at most $0.002$. A small negative bias remains at large $H$; we attribute it to the lognormal approximation of the record deviations and to the upper end of the parameter range, which truncates flat likelihoods from above. At settings where the log fit never fails, the two fits agree to within $3\%$ in RMSE, so we recommend the count fit as the default, with all-zero records and wide intervals reported alongside its RMSE.

\begin{table}[t]
\centering\footnotesize\setlength{\tabcolsep}{3pt}
\begin{tabular}{ll ccccc ccccc}
\toprule
&& \multicolumn{5}{c}{direct readout} & \multicolumn{5}{c}{level tests}\\
\cmidrule(lr){3-7}\cmidrule(lr){8-12}
&& \multicolumn{2}{c}{log fit} & \multicolumn{3}{c}{count fit} & \multicolumn{2}{c}{log fit} & \multicolumn{3}{c}{count fit}\\
\cmidrule(lr){3-4}\cmidrule(lr){5-7}\cmidrule(lr){8-9}\cmidrule(lr){10-12}
$H$ & queries & fail & bias & bias & cover & wide & fail & bias & bias & cover & wide\\
\midrule
$0.65$ & $2^{17}$ & 1\% & $-0.010$ & $-0.006$ & 0.95 / -- & 0 & 13\% & $-0.022$ & $-0.004$ & 0.94 / 0.95 & 7\%\\
$0.8$ & $2^{17}$ & 17\% & $-0.024$ & $-0.005$ & 0.93 / 0.92 & 0.7\% & 60\% & $-0.041$ & $-0.013$ & 0.94 / 0.94 & 23\%\\
$0.8$ & $2^{20}$ & 1\% & $-0.010$ & $-0.006$ & 0.93 / -- & 0 & 13\% & $-0.021$ & $-0.007$ & 0.95 / 0.95 & 0\\
$0.9$ & $2^{17}$ & 51\% & $-0.052$ & $-0.023$ & 0.95 / 0.98 & 3\% & 83\% & $-0.064$ & $-0.027$ & 0.97 / 0.98 & 26\%\\
$0.9$ & $2^{20}$ & 14\% & $-0.022$ & $-0.015$ & 0.97 / 0.98 & 0 & 45\% & $-0.050$ & $-0.027$ & 0.94 / 0.97 & 2\%\\
$0.9$ & $2^{23}$ & 0 & $-0.009$ & $-0.012$ & 0.95 / -- & 0 & 7\% & $-0.020$ & $-0.016$ & 0.97 / 0.86 & 0\\
\bottomrule
\end{tabular}
\caption{Sparse counts in the readout experiment of Table~\ref{tab:readout}, at the settings where some level has no success on at least $1\%$ of records. Log fit: percentage of records on which it fails, and bias over the remaining records. Count fit: bias over all records, coverage of the nominal $95\%$ interval over all records and over records with at least one zero count (``--'' when there are fewer than $20$ such records), and percentage of intervals wider than $0.4$. No record had zero counts at every level.}
\label{tab:sparse}
\end{table}

\subsection{Gate-Level Verification}\label{sec:verify}

Every circuit is a gate list executed by a NumPy state-vector simulator and emitted as a Qiskit~2.x circuit (Appendix~\ref{app:code}). Maximum errors were $8\times10^{-16}$ for the lattice circuit against the filter-bank matrix (db1--db4), $2\times10^{-15}$ for unitarity, $6\times10^{-17}$ for the mirror extension, $5\times10^{-16}$ for the level tests against $E_j/\|\tilde B\|^2$, $4\times10^{-16}$ for both quadratures of the phase test, and $3\times10^{-16}$ for the shift register against the nondecimated shares. These checks use state vectors.

Two further checks concern the emitted Qiskit circuits, and they differ in kind. 
The companion notebook (Appendix~\ref{app:code}) builds the gate lists for the sym4 filter as Qiskit circuits and, for a smaller three-level circuit, compares the ancilla probability computed from the Qiskit statevector with the NumPy value; this is an exact-probability check of the emitted circuits, and it was executed. A finite-shot check, which runs the level tests on Qiskit Aer for a path of length $64$ and compares the ancilla frequencies with the exact shares, is included in the archive as \texttt{qiskit\_hadamard\_hurst.py} but was not run for this paper. The CNOT counts of Table~\ref{tab:resources} were obtained from the stated decomposition rather than from a transpiler.

\section{Discussion}\label{sec:discussion}

The wavelet transform maps the scale-energy distribution into the position of the leading one in the data register. A single multiply-controlled $Z$ implements the controlled reflection for a level, and a Hadamard test reads its share through one ancilla. The lattice isolates the filter-dependent circuit operations. The mirror extension removes the endpoint jump before either decimated or nondecimated filtering, the statistical calibration incorporates the resulting boundary covariance, and Theorem~\ref{thm:clt} shows that the fold terms do not affect the first-order distribution of the estimator.

A path concentrates energy at coarse levels. The fine-level shares used by the estimator decay like $N^{-2H}$. Direct sampling has a preparation cost proportional to $p_j^{-1}$ for fixed relative precision, and amplitude estimation with the same reflection scales as $p_j^{-1/2}$ and uses coherent access to the inverse of the full preparation circuit. The simulations make the roles of the readouts precise. When the data register can be measured with ideal readout, measuring the leading qubits directly is the efficient sampling readout, because each preparation informs every level; separate Hadamard level tests spend each preparation on one level. The level reflection earns its place as the oracle of amplitude estimation without phase estimation, which needs no controlled preparation, costs within $2\%$ of sampling per query in CNOT gates although its longest circuit applies $513$ queries in sequence at $2^{20}$ queries and $4097$ at $2^{23}$, and at large $H$ is more accurate with $2^{20}$ queries than direct readout with $2^{23}$; it is also a one-qubit readout on devices where reading many qubits reliably is the binding constraint. The result we would lead with is therefore twofold: mirroring makes readout of general compactly supported wavelets viable for fBm paths, and coherent amplitude estimation improves accuracy when the informative fine-level shares are rare.

Future work includes amplitude estimation for several shares in a common window with a shared stopping rule, joint allocation of state-preparation queries across levels, Symmlet and Coiflet experiments that test the statistical and circuit tradeoffs of the lattice construction across filter families, and multivariate records, for which wavelet eigenanalysis of operator fBm \citep{AbryBonieceDidierWendt2024} suggests readouts of level covariance matrices rather than level shares. A hardware comparison of one-ancilla tests and direct leading-bit measurement would quantify how readout error affects the two protocols; the comparisons here are for ideal queries, and a practical advantage of one-ancilla readout is not claimed.

\paragraph{Code availability.}
The scripts, the companion notebook, the random seeds and the package versions used for every table and figure are archived as version~1.0 at \url{https://github.org/BraniV/QHurst}. From the archive root, \texttt{python3 hurst.py \&\& python3 readout3.py \&\& python3 mlae.py \&\& python3 analyze3.py} regenerates Tables~\ref{tab:hurst}--\ref{tab:sparse}. The scripts are: \texttt{qwt.py} (circuits and their verification), \texttt{bench.py} (Table~\ref{tab:resources}), \texttt{boundary.py} and \texttt{hurst.py} (Tables~\ref{tab:boundary} and~\ref{tab:hurst}), \texttt{lengths.py} (Figure~\ref{fig:lengths}), \texttt{readout2.py}, \texttt{readout3.py}, \texttt{countfit2.py}, \texttt{mlae.py}, \texttt{analyze3.py} and \texttt{ischeck.py} (Tables~\ref{tab:readout} and~\ref{tab:sparse} and the importance-sampling check), and the \texttt{fig\_*.py} scripts for the figures.

\paragraph{Acknowledgments.}
Supported by the National Science Foundation under Grant No.~2515246 and by the H.~O.~Hartley Chair funds at Texas A\&M University.

\appendix
\section*{Appendix}
\section{Gate Lists and Qiskit~2.x Emitter}\label{app:code}

The core of \texttt{qwt.py} is listed below. Qubit $q$ carries index bit $b_q$; gates are tuples (kind, target, [matrix or angle,] controls), with controls as (qubit, value) pairs. \texttt{python3 qwt.py} runs the verifications reported in Section~\ref{sec:sim}.

\begin{lstlisting}
def lattice(h):                                   # E(z) = R_{p-1} Lam R_{p-2} ... Lam R_0
    g = highpass(h); p = len(h) // 2
    E = [np.array([[h[2*m], h[2*m+1]], [g[2*m], g[2*m+1]]]) for m in range(p)]
    Rs = []
    while len(E) > 1:
        r1 = np.linalg.svd(E[-1])[0][:, 1]; r2 = np.array([-r1[1], r1[0]])
        if np.linalg.norm(r2 @ E[0]) > 1e-9: r2 = -r2
        R = np.column_stack([r1, r2]); F = [R.T @ Em for Em in E]
        E = [np.vstack([F[m][0], F[m+1][1]]) for m in range(len(E) - 1)]; Rs.append(R)
    Rs.append(E[0]); return Rs[::-1]

def dec_ops(reg, ctrls):                          # |v> -> |v-1 mod 2^len(reg)>
    return [("x", reg[q], ctrls + [(reg[r], 0) for r in range(q)]) for q in range(len(reg)-1, -1, -1)]
def inc_ops(reg, ctrls):                          # |v> -> |v+1 mod 2^len(reg)>
    return [("x", reg[q], ctrls + [(reg[r], 1) for r in range(q)]) for q in range(len(reg)-1, -1, -1)]

def qwt_ops(data, L, h):                          # Mallat order, finest level J-1 first
    J = len(data); Rs = lattice(h); ops = []
    for j in range(J - 1, J - L - 1, -1):
        ctrls = [(data[q], 0) for q in range(j + 1, J)]; kreg = [data[q] for q in range(1, j + 1)]
        for m, R in enumerate(Rs):
            if m > 0: ops += dec_ops(kreg, ctrls + [(data[0], 1)])   # delay: advance odd channel
            ops.append(("u", data[0], R, ctrls))
        for q in range(j): ops.append(("swap", data[q], data[q + 1], ctrls))
    return ops

def mirror_ops(data, e):                          # |0>|x> + |1>|reversed x>, e = new top qubit
    Hm = np.array([[1, 1], [1, -1]]) / np.sqrt(2)
    return [("u", e, Hm, [])] + [("x", q, [(e, 1)]) for q in data]

def shift_ops(data, sreg):                        # |s>|k> -> |s>|k+s mod N>
    return [op for b, sq in enumerate(sreg) for op in inc_ops(data[b:], [(sq, 1)])]

def level_reflection(anc, data, j):               # controlled (I - 2 P_j)
    J = len(data); return [("z", data[j], [(anc, 1)] + [(data[q], 0) for q in range(j + 1, J)])]
def threshold_reflection(anc, data, c):           # controlled (I - 2 Pi_c)
    J = len(data); return [("z", anc, [(data[J - i], 0) for i in range(1, c + 1)])]
def phase_ops(anc, data, levels, omega):          # controlled sum_j e^{i omega j} P_j + rest
    J = len(data)
    return [("p", data[j], omega * j, [(anc, 1)] + [(data[q], 0) for q in range(j + 1, J)]) for j in levels]
def hadamard_test(anc, cU, quad="re"):
    Hm = np.array([[1, 1], [1, -1]]) / np.sqrt(2)
    ops = [("u", anc, Hm, [])] + cU
    if quad == "im": ops.append(("u", anc, np.diag([1, -1j]), []))
    return ops + [("u", anc, Hm, [])]

# Level test for level j:  prep(B) on qubits 0..J-2, mirror_ops(range(J-1), J-1),
#   [shift_ops], qwt_ops(range(J), L, h), hadamard_test(anc, level_reflection(anc, range(J), j)),
#   measure anc.  P(anc = 1) = p_j.
\end{lstlisting}

The emitter \texttt{to\_qiskit} maps \texttt{"u"} to \texttt{UnitaryGate}, \texttt{"p"} to \texttt{PhaseGate}, \texttt{"x"}, \texttt{"z"} and \texttt{"swap"} to the standard gates, and controls to \texttt{.control(k, ctrl\_state=...)}. The record is loaded with \texttt{StatePreparation(B/np.linalg.norm(B))} on qubits $0,\dots,J-2$; amplitude estimation additionally needs its inverse.

The companion notebook \texttt{QuantumHurst\_QNDWT\_sym4\_Qiskit.ipynb}, included in the archive, implements the gate-list construction of this appendix for the sym4 filter, a filter outside the Daubechies family used in Section~\ref{sec:sim}, which exercises the lattice compilation for a general filter. It uses the same lists for NumPy statevector simulation and Qiskit circuit generation, estimates the Hurst exponent from ancilla counts sampled from the exact probabilities, and checks the ancilla probability against the Qiskit statevector for a smaller three-level circuit. It does not perform a finite-shot circuit simulation; that check is the unrun Aer script described in Section~\ref{sec:verify}.

\end{document}